\documentclass{PoS}\title{Improved QCD Sum-Rule Approach to
Heavy-Meson Decay Constants}\ShortTitle{Improved QCD Sum-Rule
Approach to Heavy-Meson Decay Constants}\author{Wolfgang
Lucha\\Institute for High Energy Physics, Austrian Academy of
Sciences, Nikolsdorfergasse 18, A-1050 Vienna, Austria\\E-mail:
\email{Wolfgang.Lucha@oeaw.ac.at}}\author{\speaker{Dmitri
Melikhov}\\Institute for High Energy Physics, Austrian Academy of
Sciences, Nikolsdorfergasse 18, A-1050 Vienna, Austria,\\Faculty
of Physics, University of Vienna, Boltzmanngasse 5, A-1090 Vienna,
Austria, and\\D.~V.~Skobeltsyn Institute of Nuclear Physics,
Moscow State University, 119991, Moscow, Russia\\E-mail:
\email{dmitri\_melikhov@gmx.de}}\author{Silvano Simula\\INFN,
Sezione di Roma Tre, Via della Vasca Navale 84, I-00146 Roma,
Italy\\E-mail: \email{simula@roma3.infn.it}}

\abstract{A visible improvement of the QCD sum-rule framework,
achieved only recently by a slightly more sophisticated
consideration of the hadron excitations and continuum, is applied
to the extraction of the decay constants of heavy-light
pseudoscalar mesons, such as the $D,$ $D_s,$ $B$ and $B_s$ mesons,
from two-point correlators of heavy-light pseudoscalar currents
\cite{lms2011}. This analysis is carried out entirely in terms of
the running heavy-quark mass, for which the perturbative expansion
exhibits a passable convergence. Our main concern is to gain
control over both statistical {\em and\/} systematic uncertainties
of the extracted decay constants, the former being induced by the
QCD parameter values requested for input, the latter arising from
the intrinsically limited accuracy of the QCD~sum-rule
techniques.}

\FullConference{The 2011 Europhysics Conference on High Energy
Physics, EPS-HEP 2011,\\July 21--27, 2011\\Grenoble,
Rh\^one-Alpes, France}

\begin{document}\section{Introduction}Although the QCD sum-rule
framework has been proposed even more than thirty years ago
\cite{svz}, within this approach any study of the decay constant
$f_P$ of a ground-state heavy pseudoscalar meson
$P=(Q\overline{q})$ of mass $M_P,$ viewed as bound state of a
heavy quark $Q$ and a light antiquark $\overline{q}$ with masses
$m_Q$ and $m$, respectively, still forms a nontrivial problem,
for, at least, two easily~identifiable~reasons:\begin{itemize}
\item A trustable operator product expansion (OPE) for the vacuum
two-point correlation function,\begin{equation}\Pi(p^2)\equiv{\rm
i}\int{\rm d}^4x\,\exp({\rm i}\,p\,x)\left\langle0\left|{\rm
T}\!\left(j_5(x)\,j^\dag_5(0)\right)\right|0\right\rangle,
\label{Eq:corr}\end{equation}of two pseudoscalar heavy-light
currents $j_5(x)=(m_Q+m)\,\bar q(x)\,{\rm i}\,\gamma_5\,Q(x)$ must
be constructed.\item For obvious practical reasons, the OPE is
known only in some truncated form; this means that even if its
parameters were known precisely, this limitation prevents to
extract the bound-state features with arbitrarily high accuracy
and betrays an intrinsic uncertainty of QCD
sum~rules.\end{itemize}A crucial result is that assuming
quark--hadron duality relates the lowest bound-state contribution
to the OPE for a Borel-transformed correlator $\Pi_{\rm
dual}(\tau,s_{\rm eff})$ cut at an effective continuum
threshold~$s_{\rm eff}$:\begin{equation}
\label{SR_QCD}f_P^2\,M_P^4\,\exp(-M_P^2\,\tau)=\Pi_{\rm
dual}(\tau,s_{\rm eff})\equiv\int\limits_{(m_Q+m)^2}^{s_{\rm
eff}}\hspace{-2ex}{\rm d}s\,\exp(-s\,\tau)\,\rho_{\rm
pert}(s)+\Pi_{\rm power}(\tau)\ ,\end{equation}with perturbative
spectral density $\rho_{\rm pert}(s),$ expressible as series
expansion in the strong coupling $\alpha_{\rm s}$. Before being
able to extract hadron features, we have to know the effective
continuum threshold $s_{\rm eff}$. The decay constant $f_P$
extracted from (\ref{SR_QCD}) turns out to depend on the
unphysical Borel parameter~$\tau$; the standard --- but, from our
point of view, maybe too na\"ive --- guess for a, by
assumption~constant, $s_{\rm eff}$ is found by minimizing the
$\tau$-dependence of $f_P.$ However, an analogy to
quantum-mechanics \cite{lms_qm} reveals that this does not
guarantee the reliability of the $f_P$ prediction. The reason for
this fact is that any {\em exact\/} effective continuum threshold,
found by solving Eq. (\ref{SR_QCD}) for known values of $M_P$ and
$f_P,$ will, in general, depend on $\tau.$ The requirements of
constancy of $s_{\rm eff}$ and maximal $\tau$-independence~of
$f_P$ are mutually exclusive. (In quantum mechanics, the true
bound-state masses and decay constants are found from the
solutions of the Schr\"odinger equation.) Of course, the {\em
exact\/} effective continuum threshold is not known. Thus, any
sum-rule extraction of hadronic features consists in attempting to
arrive somehow at a reasonable approximation to the true effective
continuum threshold, and to gain control over the accuracy of this
approximation; the relevant techniques have been developed~in
\cite{lms_new}.

A brief glance at the shape of the hadronic ground-state
contribution on the left-hand side of the sum rule (\ref{SR_QCD})
prompts us to introduce a dual invariant mass $M_{\rm dual}$ and a
dual decay constant~$f_{\rm dual}$~by\begin{equation}
\label{mfdual}M_{\rm dual}^2(\tau)\equiv-\frac{{\rm d}}{{\rm
d}\tau}\log\Pi_{\rm dual}(\tau,s_{\rm eff}(\tau))\ ,\qquad f_{\rm
dual}^2(\tau)\equiv M_P^{-4}\exp(M_P^2\,\tau)\,\Pi_{\rm
dual}(\tau,s_{\rm eff}(\tau))\ .\end{equation}For known true
ground-state mass, the deviation of the dual mass $M_{\rm dual}$
from the actual ground-state mass $M_P$ must be considered as
indication of the amount of excited-state contributions
picked~up~by the dual correlator defined by Eq.~(\ref{SR_QCD}).
Assuming for the effective continuum threshold a particular
functional form and requiring least deviation of the dual mass
(\ref{mfdual}) from its actual value in the range of admissible
$\tau$-values yields a variational solution for the effective
threshold. With the latter fixed, the decay constant follows from
(\ref{mfdual}). Allowing for this $\tau$-dependence of our
threshold facilitates to reproduce the actual mass and improves
the accuracy of extracted hadron observables considerably.

\section{Operator Product Expansion and Heavy-Quark Mass Scheme}
\begin{figure}[!h]\begin{center}\begin{tabular}{c}
\includegraphics[width=7.3098cm]{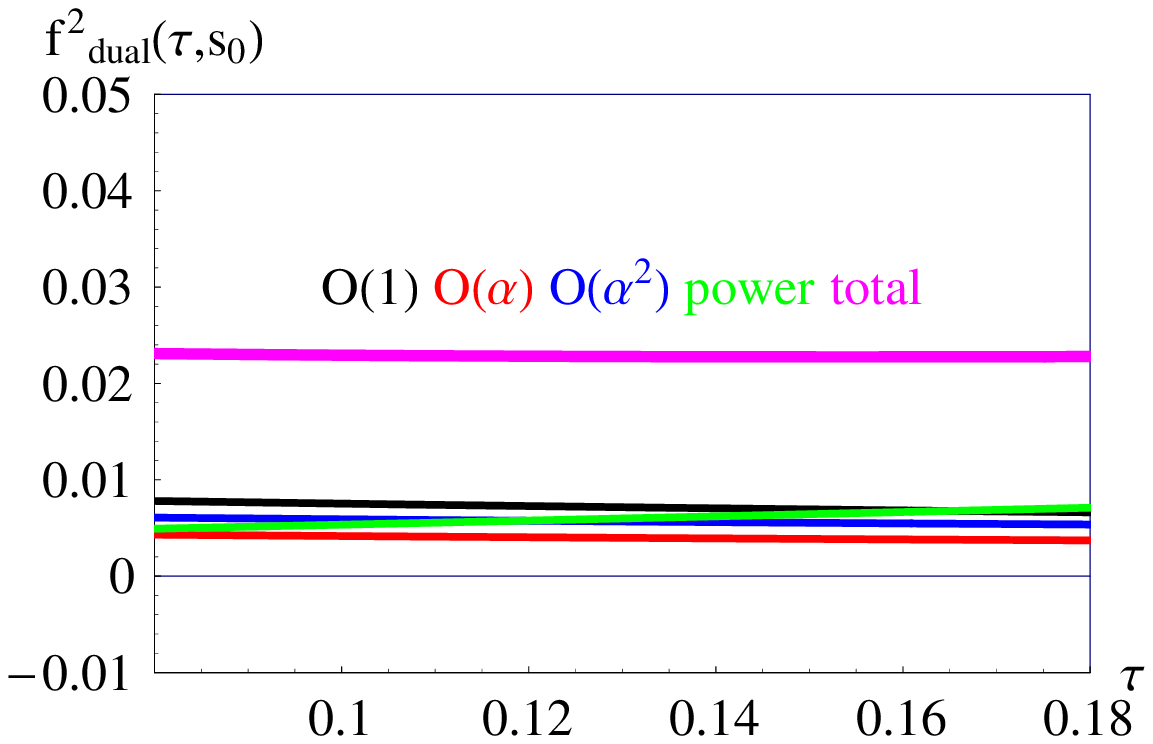}
\includegraphics[width=7.3098cm]{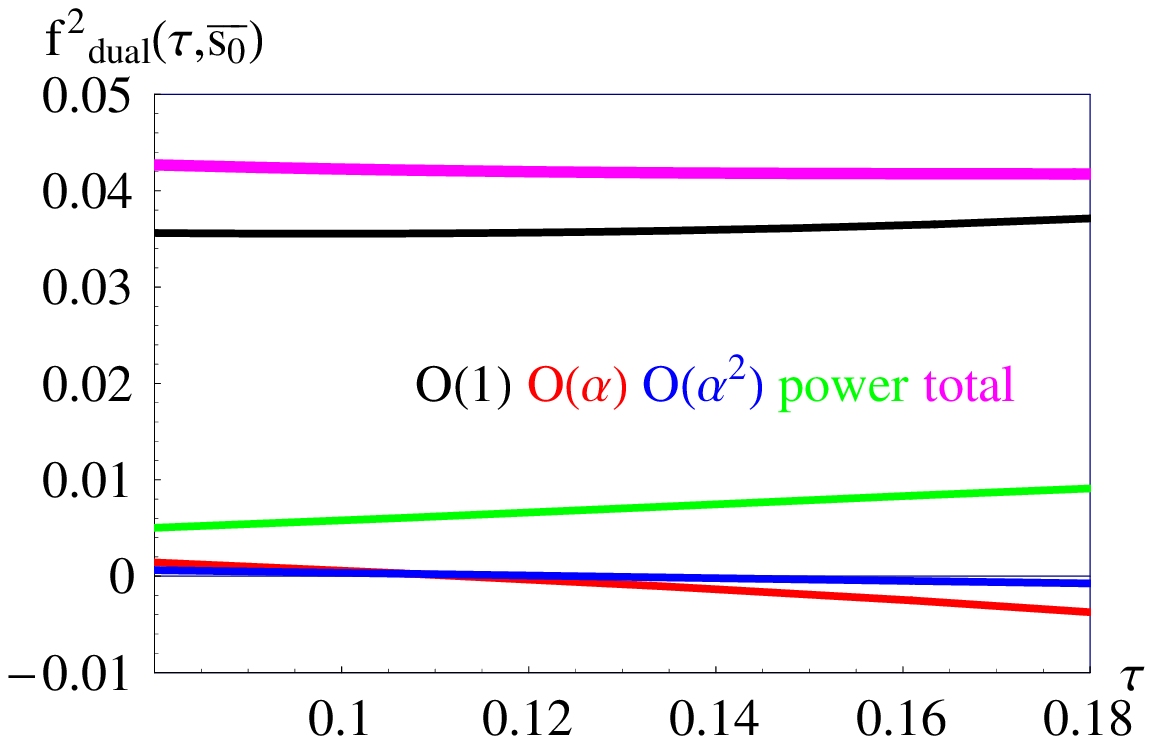}\end{tabular}
\caption{\label{Plot:1}Dual estimates $f_{\rm dual}$ for the
$B$-meson decay constant $f_B$ extracted from the OPE expressed in
terms of pole mass (left) and $\overline{\rm MS}$ mass (right) of
the $b$ quark. By requesting maximal stability of the extracted
$f_{\rm dual},$ in each case a constant effective threshold
$\stackrel{\mbox{{\tiny(}{\scriptsize\bf---}{\tiny )}}}{s_0}$ is
determined individually, thus $s_0$ and $\overline{s_0}$ differ
from~each~other. The relative power-correction contributions rise
with $\tau.$ All quantities in units of appropriate powers of
GeV.}\end{center}\end{figure}

The
choice of the precise scheme employed for the definition of the
masses of the heavy~quarks is crucial in analyses of the present
kind. The correlator (\ref{Eq:corr}) has been derived to
three-loop order in terms of the heavy quark's pole mass
\cite{chetyrkin}, which has been standard for a long time
\cite{aliev}. Alternatively, the perturbative expansion may be
reorganized in terms of the running $\overline{\rm MS}$ mass
\cite{jamin}. Figure \ref{Plot:1} shows our dual sum-rule
estimates for $f_B$ resulting for both cases. These plots allow
for important insights:\begin{enumerate}\item The perturbative
expansion for $f_{\rm dual}$ in terms of the pole mass reveals no
sign of convergence: LO, NLO, and NNLO terms contribute with
similar size and higher orders cannot be expected to give smaller
contributions. Therefore, the pole-mass OPE considerably
underestimates $f_B.$\item As already noticed in \cite{jamin},
reorganizing the perturbative expansion in terms of the running
$\overline{\rm MS}$ mass of the heavy quark yields a distinct
hierarchy of the perturbative contributions. It should be no
surprise that we prefer the OPE in terms of the $\overline{\rm
MS}$ mass in our decay-constant~analysis.\item The magnitude of
$f_{\rm dual}$ extracted from the pole-mass OPE is almost 50\%
smaller than that~one found within the $\overline{\rm MS}$ scheme;
nevertheless, in both cases the decay constants exhibit a perfect
stability in a wide range of the Borel parameter $\tau.$ Thus, we
conclude that mere Borel stability does not suffice to guarantee
the reliability of any sum-rule extraction of bound-state
features!\end{enumerate}

\section{Decay Constants of Open-Charm $D$ and $D_s$ Mesons}Our
analysis of the decay constants of the $D$ mesons \cite{lms2011}
clearly demonstrates that the improved QCD sum-rule approach
\cite{lms_new} relying on effective continuum thresholds which we
allow to depend on the Borel parameter $\tau$ yields a much better
accuracy of the quark--hadron duality approximation and a distinct
improvement of the accuracy of the extracted decay constants. This
perfectly confirms our observations \cite{lms_qcdvsqm}
(respectively, conviction) that in QCD and in quantum mechanics
the procedures~of extracting bound-state characteristics are,
qualitatively as well as quantitatively, very similar to each
other. Moreover, all estimates of systematic errors given by our
algorithm prove to be quite~realistic.

\section{Decay Constants of Open-Beauty $B$ and $B_s$ Mesons and
Running $b$-Quark Mass}The predictions of QCD sum rules for the
$B$-meson decay constants $f_B$ and $f_{B_s},$ analyzed in due
detail in \cite{lms2011}, prove to be extremely sensitive to the
precise $\overline{m}_b(\overline{m}_b)$ value used as input. For
instance, the choice $\overline{m}_b(\overline{m}_b)=(4.163\pm
0.016)\;\mbox{GeV}$ \cite{mb} entails decay-constant predictions
hardly compatible with lattice findings. However, inverting the
logic by fitting our sum-rule result for $f_B$ to the average of
lattice findings yields the very precise value of the $b$-quark
mass $\overline{m}_b(\overline{m}_b)=(4.245\pm0.025)\;{\rm GeV}.$

\section{Summary, Observations, and Conclusions}\begin{enumerate}
\item The $\tau$-dependence of effective thresholds arises quite
naturally by our attempts to increase~the precision of the duality
concept; relaxing the limitation to constant thresholds distinctly
raises both the stability of the dual mass and the quality of
sum-rule predictions for decay constants.\item Our study of {\em
charmed mesons\/} clearly demonstrates that adopting
Borel-parameter-dependent thresholds significantly improves the
accuracy of extractions of decay constants by sum rules, yields
meaningful systematic errors, reduces these to the level of a few
percent, and brings the sum-rule results into perfect agreement
with the findings of both lattice QCD and experiment.\item The
decay constants of {\em beauty mesons\/} are unexpectedly
sensitive to the chosen $\overline{m}_b(\overline{m}_b)$ value. We
may, however, regard this inconvenience as a kind of serendipity
that allows us to derive a rather accurate estimate of the
$b$-quark mass by matching the QCD sum-rule prediction~for $f_B$
to the average of the corresponding lattice evaluations. The value
of $m_b$ deduced in this way is in good agreement with lattice
findings but lacks overlap with a recent accurate determination of
$m_b$ \cite{mb} (for details, consult [1]); beyond doubt, this
disconcerting puzzle has to be resolved.\end{enumerate}

\vspace{4.74ex}\noindent{\bf Acknowledgments.} D.M.\ was supported
by the Austrian Science Fund (FWF), project no.~P22843.

\end{document}